\documentclass{PoS}
\usepackage{epsfig}
\usepackage{graphics}

\title{The colour adjoint static potential from Wilson loops with generator insertions and its physical interpretation}

\ShortTitle{The colour adjoint static potential from Wilson loops with generator insertions ...}

\author{\speaker{Marc Wagner} \\
        Goethe-Universit\"at Frankfurt am Main, Institut f\"ur Theoretische Physik, \\ Max-von-Laue-Stra{\ss}e 1, D-60438 Frankfurt am Main, Germany \\
        E-mail: \email{mwagner@th.physik.uni-frankfurt.de}}

\author{Owe Philipsen \\
        Goethe-Universit\"at Frankfurt am Main, Institut f\"ur Theoretische Physik, \\ Max-von-Laue-Stra{\ss}e 1, D-60438 Frankfurt am Main, Germany \\
        E-mail: \email{philipsen@th.physik.uni-frankfurt.de}}

\abstract{
We discuss the non-perturbative computation and interpretation of a colour adjoint static potential based on Wilson loops with generator insertions. Numerical lattice results for $SU(2)$ gauge theory are presented and compared to corresponding perturbative results.
}

\FullConference{31st International Symposium on Lattice Field Theory, LATTICE 2013\\
		July 29--August 3, 2013\\
		Mainz, Germany}

\newcommand{\bfx}{{\bf x}}
\newcommand{\bfy}{{\bf y}}
\newcommand{\bfxo}{{\bf x}_0}

\begin{document}


\section{\label{SEC002}Introduction: the singlet static potential $V^1$}

The singlet static potential $V^1$ is a common and important observable in lattice gauge theory. It is the energy of a static quark $Q(\mathbf{y})$ and a static antiquark $\bar{Q}(\mathbf{x})$ in a colour singlet orientation (i.e.\ a gauge invariant orientation) as a function of the separation $r \equiv |\mathbf{x} - \mathbf{y}|$. Since the spin of a static quark is irrelevant, static quarks will be treated as spinless colour charges in the following.

To determine the singlet static potential for any gauge group $SU(N)$ one typically defines a trial state
\begin{eqnarray}
| \Phi^{1} \rangle \ \ \equiv \ \ \bar{Q}(\bfx) U(\bfx,\bfy) Q(\bfy) | 0 \rangle ,
\end{eqnarray}
containing a static quark antiquark pair at separation $r$ in a colour singlet orientation, which has been realised by the parallel transporter $U(\bfx,\bfy)$ (on a lattice a product of links). Considering the temporal correlation function of this trial state and integrating over static quark fields one obtains the well known Wilson loop $W_1$,
\begin{eqnarray}
\langle \Phi^{1}(t_2) | \Phi^{1}(t_1) \rangle \ \ = \ \ e^{-2 M \Delta t} N \Big\langle W_1(r,\Delta t) \Big\rangle \quad , \quad \Delta t \equiv t_2 - t_1 > 0 .
\end{eqnarray}
The singlet static potential $V^1 \equiv V_0^1$ can then be extracted from the asymptotic exponential behaviour of the Wilson loop,
\begin{eqnarray}
\Big\langle W_1(r,\Delta t) \Big\rangle \ \ = \ \ \sum_{n=0}^\infty c_n \exp\Big(-V_n^1(r) \Delta t\Big) \ \ \stackrel{\Delta t \rightarrow \infty}{\propto} \ \ \exp\Big(-V^1(r) \Delta t\Big) .
\end{eqnarray}


\section{The colour adjoint static potential $V^{T^a}$}

The goal of this work is to non-perturbatively compute Wilson loops with generator insertions, which have been proposed as a definition of a colour adjoint static potential $V^{T^a}$ and are used in potential Non-Relativistic QCD (cf.\ e.g.\ \cite{pnrqcd,Brambilla:2004jw}), a framework based on perturbation theory.

\begin{figure}[htb]
\begin{center}
\includegraphics[width=0.50\textwidth]{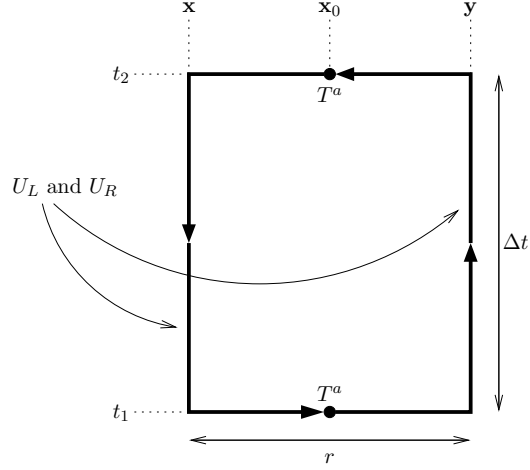}
\caption{\label{FIG001}the colour adjoint Wilson loop $W_{T^a}$.}
\end{center}
\end{figure}

A colour adjoint orientation of a static quark $Q$ and a static antiquark $\bar{Q}$, which are located at the same point in space, can be obtained by inserting one of the generators of the colour group $T^a$ (e.g.\ for $SU(3)$ one of the Gell-Mann matrices, $T^a = \lambda^a/2$), i.e.\ $\bar{Q} T^a Q$. In case the static charges are separated in space, a straightforward generalisation are trial states
\begin{eqnarray}
| \Phi^{T^a} \rangle \ \ \equiv \ \ \bar{Q}(\bfx) U(\bfx,\bfxo) T^a U(\bfxo,\bfy) Q(\bfy) | 0 \rangle .
\end{eqnarray}
In the following we discuss non-perturbative calculations of the corresponding colour adjoint static potential in various gauges analogous to that of the singlet static potential in section~\ref{SEC002},
\begin{eqnarray}
 & & \hspace{-0.7cm} \langle \Phi^{T^a}(t_2) | \Phi^{T^a}(t_1) \rangle \ \ = \ \ e^{-2 M \Delta t} N \Big\langle W_{T^a}(r,\Delta t) \Big\rangle \quad , \quad W_{T^a}(r,\Delta t) \ \ \equiv \ \ \frac{1}{N} \textrm{Tr}\Big(T^a U_R T^{a,\dagger} U_L\Big) \\
\label{EQN003} & & \hspace{-0.7cm} \Big\langle W_{T^a}(r,\Delta t) \Big\rangle \ \ = \ \ \sum_{n=0}^\infty c_n \exp\Big(-V_n^{T^a}(r) \Delta t\Big) \ \ \stackrel{\Delta t \rightarrow \infty}{\propto} \ \ \exp\Big(-V^{T^a}(r) \Delta t\Big)
\end{eqnarray}
(cf.\ also Figure~\ref{FIG001}).
In particular we are interested in,
\begin{itemize}
\item[(1)] whether the colour adjoint static potential $V^{T^a} \equiv V_0^{T^a}$ is gauge invariant (i.e.\ whether the obvious gauge dependence of the correlation function $\langle W_{T^a}(r,\Delta t) \rangle$ only appears in the matrix elements $c_n$),

\item[(2)] whether $V^{T^a}$ indeed corresponds to the potential of a static antiquark and a static quark in a colour adjoint orientation, or whether it has to be interpreted differently.
\end{itemize}

More details regarding this work can be found in \cite{Wagner:2012pk,Philipsen:2013ysa}.


\subsection{$V^{T^a}$ without gauge fixing}

Without gauge fixing
\begin{eqnarray}
\Big\langle W_{T^a}(r,\Delta t) \Big\rangle \ \ = \ \ 0 ,
\end{eqnarray}
because $W_{T^a}(r,\Delta t)$ is gauge variant and does not contain any gauge invariant contribution. Clearly, without gauge fixing the calculation of a colour adjoint static potential using (\ref{EQN003}) fails.


\subsection{$V^{T^a}$ in Coulomb gauge}

Coulomb gauge, $\nabla \mathbf{A}^g(x) = 0$, amounts to an independent condition on every time slice $t$. It is not complete. The remaining residual gauge symmetry corresponds to global independent colour rotations $h^{\scriptsize \textrm{res}}(t) \in SU(N)$ on every time slice $t$. With respect to this residual gauge symmetry the colour adjoint Wilson loop transforms as
\begin{eqnarray}
\Big\langle W_{T^a}(r,\Delta t) \Big\rangle \ \ \rightarrow_{h^{\scriptsize \textrm{res}}} \ \ \frac{1}{N} \textrm{Tr}\Big(h^{{\scriptsize \textrm{res}},\dagger}(t_1) T^a h^{\scriptsize \textrm{res}}(t_1) U_R h^{\scriptsize \textrm{res}}(t_2) T^{a,\dagger} h^{{\scriptsize \textrm{res}},\dagger}(t_2) U_L\Big) .
\end{eqnarray}
Since $h^{\scriptsize \textrm{res}}(t_1)$ and $h^{\scriptsize \textrm{res}}(t_2)$ are independent, the situation is analogous to that without gauge fixing, i.e.\
\begin{eqnarray}
\Big\langle W_{T^a}(r,\Delta t) \Big\rangle_{\scriptsize \textrm{Coulomb gauge}} \ \ = \ \ 0 .
\end{eqnarray}
Therefore, also in Coulomb gauge the non-perturbative calculation of a colour adjoint static potential fails.


\subsection{$V^{T^a}$ in Lorenz gauge}

In Lorenz gauge, $\partial_\mu A_\mu^g(x) = 0$, a Hamiltonian or a transfer matrix does not exist. Therefore, only gauge invariant correlation functions like the ordinary Wilson loop $\langle W_1(r,\Delta t) \rangle$ exhibit an asymptotic exponential behaviour and allow the determination of energy eigenvalues. The colour adjoint Wilson loop $\langle W_{T^a}(r,\Delta t) \rangle_{\scriptsize \textrm{Lorenz gauge}}$, on the other hand, does not decay exponentially in the limit of large $\Delta t$. Hence, the physical meaning of a colour adjoint static potential determined from $\langle W_{T^a}(r,\Delta t) \rangle_{\scriptsize \textrm{Lorenz gauge}}$ is unclear.


\subsection{$V^{T^a}$ in temporal gauge}

The implementation of temporal gauge in the continuum based on the Feynman propagation kernel is given in \cite{Rossi:1979jf,Rossi:1980pg}. We follow the lattice formulation based on the transfer matrix \cite{Creutz:1976ch,Philipsen:2001ip,jp}.

Temporal gauge amounts to link variables $U_0^g(x) = 1$. Temporal links gauge transform as $U_0^g(t,\mathbf{x}) = g(t,\mathbf{x}) U_0(t,\mathbf{x}) g^\dagger(t+a,\mathbf{x})$, $g(t,\mathbf{x}) \in SU(N)$. Consequently, a possible choice to implement temporal gauge is $g(t=2a,\mathbf{x}) = U_0(t=a,\mathbf{x})$, $g(t=3a,\mathbf{x}) = U_0(t=a,\mathbf{x}) U_0(t=2a,\mathbf{x})$, ...

By inserting this transformation to temporal gauge $g(t,\mathbf{x})$ the gauge variant colour adjoint Wilson loop turns into a gauge invariant observable,
\begin{eqnarray}
\nonumber & & \hspace{-0.7cm} \Big\langle W_{T^a}(r,\Delta t) \Big\rangle_{\scriptsize \textrm{temporal gauge}} \ \ = \\
\nonumber & & = \ \ \frac{1}{N} \Big\langle \textrm{Tr}\Big(U^{T^a,g}(t_1;\bfx,\bfy) U^{T^{a,\dagger},g}(t_2;\bfy,\bfx)\Big) \Big\rangle_{\scriptsize \textrm{temporal gauge}} \ \ = \ \ \ldots \ \ = \\
\label{EQN004} & & = \ \ \frac{2}{N (N^2-1)} \sum_a \sum_b \Big\langle \textrm{Tr}\Big(T^a U_R T^b U_L\Big) \textrm{Tr}\Big(T^a U(t_1,t_2;\bfxo) T^b U(t_2,t_1;\bfxo)\Big) \Big\rangle
\end{eqnarray}
($U^{T^a}(\bfx,\bfy) = U(\bfx,\bfxo) T^a U(\bfxo,\bfy)$; cf.\ \cite{Philipsen:2013ysa} for details). $\textrm{Tr}(T^a U_R T^b U_L)$ denotes a Wilson loop with generator insertions at the spatial sides, while $\textrm{Tr}(T^a U(t_1,t_2;\bfxo) T^b U(t_2,t_1;\bfxo))$ is proportional to the propagator of a static adjoint quark. Consequently, the colour adjoint Wilson loop in temporal gauge is a correlation function of a gauge invariant three-quark state, one fundamental static quark, one fundamental static anti-quark, one adjoint static quark.

Equivalently, after defining a trial state with three static quarks,
\begin{eqnarray}
| \Phi^{Q \bar{Q} Q^{\scriptsize \textrm{ad}}} \rangle \ \ \equiv \ \ Q^{{\scriptsize \textrm{ad}},a}(\bfxo) \Big(\bar{Q}(\bfx) U^{T^a}(\bfx,\bfy) Q(\bfy)\Big) |0\rangle ,
\end{eqnarray}
one can verify
\begin{eqnarray}
\langle \Phi^{Q \bar{Q} Q^{\scriptsize \textrm{ad}}}(t_2) | \Phi^{Q \bar{Q} Q^{\scriptsize \textrm{ad}}}(t_1) \rangle \ \ \propto \ \ \Big\langle W_{T^a}(r,\Delta t) \Big\rangle_{\scriptsize \textrm{temporal gauge}} .
\end{eqnarray}
The conclusion is that $V^{T^a}$ in temporal gauge should not be interpreted as the potential of a static quark and a static anti-quark, which form a colour-adjoint state, but rather as the potential of a colour-singlet three-quark state. Note that this potential does not only depend on the $Q \bar{Q}$ separation $r=|\bfx-\bfy|$, but also on the position $\bfxo$ of the generator $T^a$, which is also the position of the static adjoint quark (in the following we work with the symmetric alignment $\bfxo=(\bfx+\bfy)/2$).

Using the transfer matrix formalism yields the same result. One can perform a spectral analysis of the colour adjoint Wilson loop,
\begin{eqnarray}
\Big\langle W_{T^a}(r,\Delta t) \Big\rangle_{\scriptsize \textrm{temporal gauge}} \ \ = \ \ \frac{1}{N} \sum_{k} e^{-(V_{k}^{T^a}(r) - \mathcal{E}_0) \Delta t} \sum_{\alpha,\beta} \Big|\langle k^a_{\alpha \beta} | U^{T^a}_{\alpha \beta}(\bfx,\bfy) |0\rangle\Big|^2 ,
\end{eqnarray}
where $| k^a_{\alpha \beta} \rangle$ denotes states with two fundamental indices $\alpha$ and $\beta$ and one adjoint index $a$. In terms of colour charges this is equivalent to one fundamental static quark, one fundamental static anti-quark, and one adjoint static quark in a colour singlet orientation. Again the conclusion is that $V^{T^a}$ in temporal gauge is the potential of a colour singlet three-quark state (cf.\ \cite{Philipsen:2013ysa} for details).


\section{A gauge invariant definition of the colour adjoint static potential using $\mathbf{B}$ fields}

Another proposal from the literature (cf.\ e.g.\ \cite{pnrqcd,Brambilla:2004jw}) to determine the colour adjoint static potential is based on the gauge invariant quantity
\begin{eqnarray}
W_B(r,\Delta t) \ \ \equiv \ \ \frac{1}{N} \textrm{Tr}\Big(T^a U_R T^{b,\dagger} U_L\Big) \mathbf{B}^a(\bfxo,t_1) \mathbf{B}^b(\bfxo,t_2) ,
\end{eqnarray}
where the open colour indices of the Wilson loop with generator insertions are saturated by colour magnetic fields. Using the transfer matrix formalism one can again perform a spectral analysis and show that only states $| k_{\alpha \beta} \rangle$ from the quark antiquark colour singlet sector contribute to the correlation function (in this case with opposite parity compared to the standard singlet static potential; for a detailed discussion of quantum numbers of states with two static charges we refer to e.g.\ \cite{Bali:2005fu,Wagner:2010ad}):
\begin{eqnarray}
\Big\langle W_B(r,\Delta t) \Big\rangle \ \ = \ \ \sum_{k} e^{-(V_{k}^{1,-}(r) - \mathcal{E}_0) \Delta t} \sum_{\alpha,\beta} \Big|\langle k_{\alpha \beta} | U^{T^a \mathbf{B}^a}_{\alpha \beta}(\bfx,\bfy) |0\rangle\Big|^2 .
\end{eqnarray}
Therefore, also $\langle W_B(r,\Delta t) \rangle$ is not suited to extract a colour adjoint static potential. It can, however, be used to extract so-called hybrid potentials (cf.\ e.g.\ \cite{Bali:2000gf,Juge:2002br,Bali:2003jq}).


\section{Numerical results from $SU(2)$ lattice gauge theory}

\begin{figure}[htb]
\begin{center}
\includegraphics[angle=-90,width=0.55\textwidth]{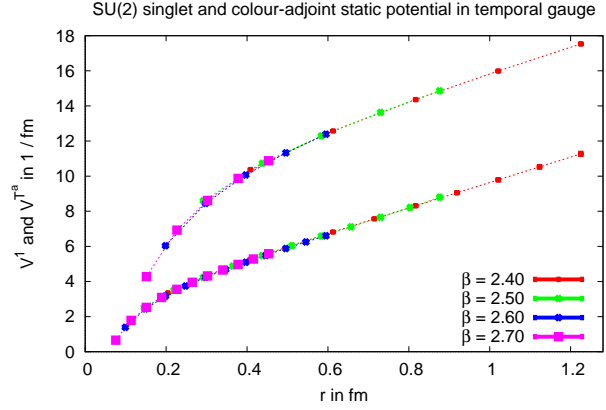}
\caption{\label{FIG002}$SU(2)$ lattice result for the singlet potential $V^1$ (lower curve) and the colour adjoint static potential $V^{T^a}$ in temporal gauge (upper curve).}
\end{center}
\end{figure}

We have performed an $SU(2)$ lattice computation of both the singlet potential $V^1$ and the colour adjoint static potential $V^{T^a}$ in temporal gauge (which should be interpreted as a $Q \bar{Q} Q^{\scriptsize \textrm{ad}}$) static potential). Results obtained at four different lattice spacings $a = 0.038 \, \textrm{fm} \ldots 0.102 \, \textrm{fm}$ are shown in Figure~\ref{FIG002}.
$V^{T^a}$ in temporal gauge is for small static quark separations even stronger attractive than the singlet static potential $V^1$. For large separations both potentials exhibit approximately the same slope, which indicates flux tube formation between $Q Q^{\scriptsize \textrm{ad}}$ and $\bar{Q} Q^{\scriptsize \textrm{ad}}$.


\section{Leading order perturbative calculations}

Perturbation theory for a static potential is a good approximation for small quark separations and should agree in that region with corresponding non-perturbative results. For the gauge invariant singlet static potential the leading order result
\begin{eqnarray}
V^1(r) \ \ = \ \ -\frac{(N^2 - 1) g^2}{8 N \pi r} + \textrm{const} + \mathcal{O}(g^4)
\end{eqnarray}
is well-known and reproduces the Coulomb-like attractive behaviour observed in numerical lattice computations at small separations (cf.\ Figure~\ref{FIG002}). The leading order colour adjoint static potential in Lorenz gauge,
\begin{eqnarray}
V^{T^a}(r) \ \ = \ \ +\frac{g^2}{8 N \pi r} + \textrm{const} + \mathcal{O}(g^4) .
\end{eqnarray}
appears frequently in the literature. Since in Lorenz gauge a Hamiltonian or a transfer matrix does not exist, its physical meaning is unclear. Moreover, its Coulomb-like repulsive behaviour is not reproduced by any of the previously presented non-perturbative considerations or computations. To study the colour adjoint static potential in temporal gauge perturbatively, we have calculated the leading order of the equivalent gauge invariant expression derived in (\ref{EQN004}) in Lorenz gauge:
\begin{eqnarray}
\nonumber & & \hspace{-0.7cm} V^{T^a}(r,\bfxo=(\bfx+\bfy)/2)\Big|_{\scriptsize \textrm{temporal gauge}} \ \ = \ \ V^{Q \bar{Q} Q^{\scriptsize \textrm{ad}}}(r,\bfxo=(\bfx+\bfy)/2) \ \ = \\
 & & = \ \ -\frac{(4 N^2 - 1) g^2}{8 N \pi r} + \textrm{const} + \mathcal{O}(g^4) .
\end{eqnarray}
It is attractive, stronger by a factor $4 \ldots 5$ than the singlet static potential (depending on $N$), which is in agreement with numerical lattice results (cf.\ Figure~\ref{FIG002}).


\section{Conclusions}

We have discussed the non-perturbative definition of a static potential for a quark antiquark pair in a colour adjoint orientation, based on Wilson loops with generator insertions $W_{T^a}(r,\Delta t)$ in various gauges:
\begin{itemize}
\item \textbf{Without gauge fixing, Coulomb gauge:} $\langle W_{T^a}(r,\Delta t) \rangle = 0$, i.e.\ the calculation of a potential $V^{T^a}$ fails.

\item \textbf{Lorenz gauge:} a Hamiltonian or a transfer matrix does not exist, the physical meaning of a corresponding potential $V^{T^a}$ is unclear.

\item \textbf{Temporal gauge:} a strongly attractive potential $V^{T^a}$, which should be interpreted as the potential of three quarks, i.e.\ $V^{T^a} = V^{Q \bar{Q} Q^\textrm{ad}}$.
\end{itemize}
When saturating open colour indices with colour magnetic fields $\mathbf{B}^a$, one obtains a singlet static potential.

On the other hand leading order perturbative calculations in Lorenz gauge have long predicted $V^{T^a}$ to be repulsive. It appears impossible, to reproduce this repulsive behaviour by a non-perturbative computation based on Wilson loops with generator insertions.

In a recent similar work a non-perturbative extraction of the colour-adjoint potential from Polyakov loop correlators was suggested \cite{Rossi:2013qba,Rossi:2013iba}. Similar to our treatment here and in earlier work \cite{Wagner:2012pk,Philipsen:2013ysa}, an adjoint Schwinger line appears, which however is placed at spatial infinity, or far away from the fundamental quarks. While no simulations of this observable are available yet, since adjoint charges are screened we expect this correlator to decay as the ordinary Polyakov loop correlator with the singlet potential shifted by a gluelump mass.


\begin{acknowledgments}

We thank Felix Karbstein and Antonio Pineda for discussions. M.W.\ acknowledges support by the Emmy Noether Programme of the DFG (German Research Foundation), grant WA 3000/1-1. This work was supported in part by the Helmholtz International Center for FAIR within the framework of the LOEWE program launched by the State of Hesse.

\end{acknowledgments}



\end{document}